\documentclass[a4paper, 10pt]{scrartcl}

\usepackage{amsfonts}
\usepackage[figuresright]{rotating}  
\usepackage{amssymb}
\usepackage{amsmath}
\usepackage{bm}
\usepackage{a4wide}

\usepackage{graphicx}
\usepackage[super]{cite}

\usepackage{paralist}
\usepackage[caption=false]{subfig}
\usepackage{pinlabel}
\usepackage{url}
\usepackage{booktabs}
\usepackage[capitalise]{cleveref}

\newcommand*{\ee}{\ensuremath{\mathrm{e}}}
\newcommand*{\ii}{\ensuremath{\mathrm{i}}}
\newcommand*{\dd}{\ensuremath{\,\mathrm{d}}}
\newcommand*{\vecc}[1]{\ensuremath{\bm{#1}}}

\newcommand*{\vx}{\vecc{x}}
\newcommand*{\vxi}{\vecc{\xi}}
\newcommand{\norm}[1]{\left\Vert#1\right\Vert}
\newcommand{\abs}[1]{\left\vert#1\right\vert}

\newcommand{\Real}{\mathbb R}

\DeclareFontFamily{U}{wncy}{}
\DeclareFontShape{U}{wncy}{m}{n}{<->wncyr10}{}
\DeclareSymbolFont{mcy}{U}{wncy}{m}{n}
\DeclareMathSymbol{\Sh}{\mathord}{mcy}{"58} 

\newcommand*\colvec[3][]{
	\begin{pmatrix}\ifx\relax#1\relax\else#1\\\fi#2\\#3\end{pmatrix}
}
\def\Pr{\mathop{\mathrm P}\nolimits}

\begin{document}

\markboth{Oleg Soloviev}{Alias-free basis for sensorless adaptive optics}

\title{Alias-Free Basis for Modal Sensorless Adaptive Optics Using the Second Moment of Intensity
}

\author{Oleg Soloviev\footnote{DCSC, 3mE, Delft University of Technology,
		Mekelweg 2, 2628 CD Delft, the Netherands,
			o.a.soloviev@tudelft.nl} 
}



\maketitle

\begin{abstract}
In theory of optical aberrations, an aberrated wavefront is represented by its coefficients in some orthogonal basis, for instance by Zernike polynomials.
However, many wavefront measurement techniques implicitly approximate the gradient of the wavefront by the gradients of the basis functions.
For a finite number of approximation terms, the transition from a basis to its gradient might introduce an aliasing error.
To simplify the measurements, another set of  functions, an ``optimal basis'' with orthogonal gradients, is often introduced, for instance Lukosz-Braat polynomials.
The article first shows that such  bases do not necessarily eliminate the aliasing error and secondly considers the problem of finding an alias-free basis on example of second-moment based indirect wavefront sensing methods.
It demonstrates that for these methods any alias-free basis should be formed by functions simultaneously orthogonal in two dot-products and be composed of the eigenfunctions of the Laplace operator.
The fitness of such alias-free basis for optical applications is analysed by means of numerical simulations on typical aberrations occurring in microscopy and astronomy.

\end{abstract}

\section{Introduction}

In   optics, the quality of an optical system can be characterised by the \emph{rms} of the  wavefront aberration.
Thus, in adaptive optics, the \emph{rms} of the \emph{residual} wavefront error is the measure of the correction:
\begin{equation}\label{eq: error}
\epsilon = \int_A (\phi(\vxi) - \varphi_{DM}(\vxi))^2  \dd \vxi,
\end{equation}
where $\phi$ and $\varphi_{DM}$ denote the incoming and correction wavefronts, and $\vxi$ is  the coordinate vector inside  aperture $A$.
On the other hand, wavefront sensors often provide information based on the gradient of the wavefronts, and control algorithms  use a quantity related (maybe implicitly) to the \emph{rms} of the gradient of residual aberration:
\begin{equation}\label{eq: grad error}
\epsilon_{\nabla} = \int_A \norm{\nabla\phi(\vxi) - \nabla\varphi_{DM}(\vxi)}^2  \dd \vxi.
\end{equation}
While without limitation on $\varphi_{DM}$, minimisation of both errors  gives equivalent up to a constant phase term results, this is not the case when $\varphi_{DM}$ is confined to some subspace, for instance, spanned by the response functions of a deformable mirror.

This article considers an abstract mathematical problem of finding such a basis that minimisation of both error given by \cref{eq: error,eq: grad error} over any subspace spanned by a finite subset of the basis functions provides equivalent results and discusses the advantages of using such bases in optical applications on example of indirect wavefront sensing method based on the second moment of intensity.
The method is briefly recapitulated in the following section.

\section{Indirect wavefront sensing}


In feedback adaptive optics (AO) systems, the goal of wavefront correction is to minimise the \emph{rms} of the residual aberration~\cite{Roddier1999},
which is measured by means of the wavefront sensing.
In applications with (quasi-)static aberrations, such as microscopy, indirect (also referred to as wavefront sensor-less) methods provide some advantage in comparison with direct wavefront sensing and at present can be considered as a common approach to aberration sensing and correction in AO microscopy~\cite{Booth2014}.
In these methods, the aberration is corrected by optimising some image quality metric as a function of the control signals applied to an adaptive element.
The chosen metric should  of course be related to the \emph{rms} value of the residual aberration, so that optimisation of the metrics should minimise the wavefront error. 

In model-based indirect methods,  the correction wavefront is modelled as a linear  function of some parameters $\vecc{c} = (c_1,\ldots,c_N)^T$, where $N$ is the number of the degrees of freedom in the model;
the parameters  $\vecc{c}$ are often (but not necessarily) linearly related to the control signals $\vecc{v}$ of the adaptive element: $\vecc{v} = \vecc{L} \vecc{c}$.
In this case one effectively controls the adaptive element via modes formed from the response functions by columns of matrix $\vecc{L}.$

If the image quality  metric $J$ attains its optimum value $J_o$ at some optimal point $\vecc{c}_o = (c_{1,o},\ldots,c_{N,o})^T$,
it  can be considered as quadratic in the vicinity of $\vecc{c}_o$: 
\begin{equation}\label{eq: quadratic form}
J(c_1,\ldots,c_N) \approx J_o \pm \sum_{i,j =1}^N m_{ij} (c_i-c_{i,o})(c_j - c_{j,o}), \quad
\text{provided} \norm{\vecc{c}-\vecc{c}_o} \text {is small},
\end{equation} 
where $m_{ij}$ are the elements of some positive semidefinite matrix $\vecc{M}\succeq 0$.
In this case, the metric can be optimised using one or more iterations consisting of $P N +1$ independent images each~\cite{Facomprez2012}, where $P$ is some integer number, provided the coefficients $m_{ij}$ of the quadratic form are known.

The structure of matrix $\vecc{M}$ is influenced by the choice of the metric and 
the choice of the modes (that is by defining matrix $\vecc{L}$).
The particular case of the  diagonal  matrix is obviously the most advantageous, as it allows to adjust each of the modes independently, which increases the accuracy of the wavefront reconstruction and decreases the number of required iterations; it is achieved by careful selection of the metric function, or by selection of the modes that diagonalises the matrix $m_{ij}$:
\begin{equation}\label{eq:m diagonalisation}
\vecc{M} = \vecc{V}\vecc{S} \vecc{V}^T,
\end{equation}
where $\vecc{S}$ is some positive semidefinite diagonal matrix and the rows of $\vecc{V}$ define the new modes.
The shape and properties of $\vecc{V}$ depend on the orthogonalisation procedure used: 
for instance, Gram-Schmidt orthogonalisation returns a  lower-triangular $\vecc{V}$; and in SVD  $\vecc{V}$ is an orthogonal matrix formed by the eigenvectors of $\vecc{M},$ with $\vecc{S}$ is formed by the eigenvalues of $\vecc{M}$.

The modal functions that diagonalise the metric $J$  are  usually referred to as ``optimal''~\cite{Debarre:07,Wang2009,Turaga2010,Thayil2011} and can be obtained analytically, numerically  or empirically~\cite{Thayil2011}. 

A special class of metrics is based on measuring the change of the second moment (SM) of the image.
This method is applicable both to an image of a point source~\cite{Booth2007,Linhai2011} and of an extended object~\cite{Yang2015}. This metric can be expressed via the squared gradient of the residual aberration; consequently it is quadratic also for larger aberrations and can be optimised in $N+1$ steps.

It has been shown that the Lukosz-Braat (LB) polynomials are the optimal modes for the second-moment methods~\cite{Booth2007,Debarre:07}.
The LB polynomials are obtained from Zernike polynomials by Gram-Schmidt  orthogonalisation of their gradients~\cite{Braat1987} and thus minimise the transverse ray aberration (the second moment of a PSF). This is an example of an analytically obtained optimal basis.

For a low-order deformable mirror, the LB polynomials cannot be approximated well enough with the mirror response functions, and  the optimal modes are obtained by SVD orthogonalisation of their gradients~\cite{Wang2009}, or by SVD decomposition of an empirically obtained matrix $\vecc{M}$ using \cref{eq:m diagonalisation}. Modes obtained in such a way can be considered as numerical and empirical optimal bases respectively; this approach is more often used in practice. It was shown~\cite{Wang2009} that use of SVD modes is more advantageous over approximated LB modes for an adaptive element with a small number of actuators.

The subject of this paper is the derivation of an 
analytical 
optimal basis for the case when the number of degrees of freedom of an adaptive element is very large, such as for SLM or photo-addressed DM~\cite{Bonora2012}. 
By ``very large'' or ``practically infinite'' we assume that the adaptive element is able to represent any analytical mode of interest accurately enough. 
We can mark the following differences with the case of a small number of degrees of freedom.

Firstly, because of the increased dimensionality, it becomes unpractical to find  the optimal modes via SVD  and they should be determined analytically --- one needs to perform $N^2$ measurements to empirically obtain matrix $\vecc{M}$, and the results of this calibration procedure would be prone to errors, as a higher density of actuators usually results in decreased amplitude of individual response functions.

Secondly, unlike the small-dimensional case, where all optimal bases are related to each other via a (small-dimensional) isometric transformation~\cite{Debarre2009}, in a large-dimensional case, the corresponding transformation would result in a computation of large series, prone to the numerical errors; and anyway this procedure is effectively equivalent to just defining  new modes point-wise.

At last, while in the case of a small number of actuators one can use all of them to obtain the best correction that can be achieved, in big-dimensional situations one needs to limit effective number of degrees of freedom to some smaller value  $N'$ 
in order to avoid overexposure of the specimen;
in general, one wants to keep $N'$ as small as possible. 
It is natural to think that  any incoming aberration $\phi$  can be decomposed as a sum of two  functions, one from the space that an adaptive element (or only those its modes that are used) can correct and the residual which cannot be corrected, so one can  simply ignore it:
\begin{equation}\label{eq: DM and not DM functions}
\phi = \sum_{n=1}^{N'} f_n +\sum_{n>N'} f_n = \varphi_{DM} + \varphi_r,
\end{equation}
where, to make the decomposition unique, the function $\varphi_r$ is chosen in the least square-sense approximation, which is equivalent to the condition that $\varphi_{DM}$ and $\varphi_r$ are mutually orthogonal.
Then, following this reasoning, matrix $\vecc{M}$ is obtained and diagonalised only for the first $N'$ basis terms.
However, it is often forgotten that there are functions that cannot be corrected but which can still  be ``seen'' by the metric and, consequently, there are functions with a non-zero correctable part $\varphi_{DM}$ that will not be seen by the method and remain under-corrected.
In this situation, the accuracy of the correction using only the first $N'$ modes would not only depend on the statistical properties of the aberration, but also on the way the optimal basis was obtained. 
For instance, LB modes, obtained by sequential orthogonalisation of the gradients of the Zernike modes by Gram-Schmidt algorithm, introduce a systematic error  as is shown in Section~\ref{seq: aliasing}.

The rest of the article is organised as follows. 
First, we show on an example the problem of aliasing if suboptimal modes are used. 
Second, we show that second-moment metric infers a new dot product in the space of modes and show that diagonalisation of $\vecc{M}$ is equivalent to orthogonalisation of the modal basis. 
Then  we derive the condition for the aliasing-free basis. 
Finally, we present and discuss the results of numerical simulations using optimal modes on examples of typical aberration in microscopy and in astronomy.

\section{Aliasing error in the second-moment methods}

In this section we demonstrate how use of the first terms of a basis, although optimal in the sense of diagonalisation of matrix $\vecc{M}$, can result in a  systematic aliasing error. 
Although the aliasing or coupling error is familiar to all using adaptive optics with direct or indirect wavefront sensing, we dwell a little bit on it to demonstrate that the error is caused completely by the choice of basis and can be avoided by using an ``even more optimal'' basis.
We name the bases correspondingly as \emph{suboptimal} and \emph{optimal} (or \emph{aliasing-free}) bases.

In the examples, we use indirect wavefront sensing based on  any image metric related to the computation of the second moment of the intensity in the focal plane. 
The concise description of such a metric for images of a point source is given in the next section.
The same reasoning, however, can be applied and similar examples can be constructed for other metrics.

\subsection{Second-moment based wavefront sensor-less method}
For an image of a point source obtained with control signals $\vecc{v}$,  let metric $J(\vecc{v})$ be defined as a second moment of its normalised intensity  $I(x,y) = I(x,y|\vecc{v})$:
\begin{equation}\label{eq: sm discrete}
J(\vecc{v}) = \frac{\sum_{x,y} I(x,y) (x^2 + y^2)}{\sum_{x,y} I(x,y)} ,
\end{equation}
where the origin $(x,y) = 0$ is given by the intersection of the optical axis with the imaging plane.

Under some natural limitations, the sums in the equation above can be considered as discrete approximations of two-dimensional integrals over $\mathbb{R}^2$, and by denoting as $\varphi = \varphi(\vecc{v})$ and $\phi$ the phase  generated by the adaptive element and  that of the initial aberration, in the case of uniform illumination, metric $J$ can be shown~\cite{Yang2015} to be linear dependent on the averaged square gradient of the phase of light field in the aperture:
\begin{equation}\label{eq: SM integral on v}
J(\vecc{v}) \approx J_d + 
\frac{1}{4 \pi^ 2 } 
\frac{ \int_{\Real^2}  {P^2(\vxi)}  \abs{\nabla (\phi- \varphi(\vecc{v}))}^2 \dd \vxi}{ \int_{\Real^2} P^2(\vxi) \dd \vxi}
,
\end{equation}
where $\vxi = (\xi, \eta)$ is the coordinate vector in the pupil plane, $P(\vxi)$ is the pupil function, and $J_d$ is the second moment of the diffraction-limited PSF.

Assuming linear dependence of the correction phase on the control signals
\begin{equation}
\varphi (\vecc{v}) = \sum_{n=1}^{N} f_n(\vxi) v_n,
\end{equation}
where $f_n(\vxi), \ n =1, \ldots, N$ are the response functions of the adaptive element,  expanding the modulus square in the right hand side of ~\cref{eq: SM integral on v} and completing the full square, the metric $J$ can now be written in the form of \cref{eq: quadratic form} as 
\begin{equation}\label{eq: full_square}
J(\vecc{v}) \approx J_d + m_o + (\vecc{v} -\vecc{v}_o)^T \vecc{M} (\vecc{v} -\vecc{v}_o),
\end{equation}
where $m_o \geq 0$ is some constant dependent on $\varphi$, such that the minimum value $J_o$ of \cref{eq: quadratic form} is given by  $J_o = J_d + m_o$, and the elements $m_{i,j}$ of the matrix $\vecc{M}$ are given by the averaged inner product of the gradients of the response functions $f_i$ and $f_j$:
\begin{equation}\label{eq: sm matrix}
m_{i,j} = \frac{1}{4 \pi^ 2 } 
\frac{ \int_{\Real^2}  {P^2(\vxi)}  (\nabla f_i, \nabla f_j)  \dd \vxi}{ \int_{\Real^2} P^2(\vxi) \dd \vxi}.
\end{equation}
If all  elements $m_{i,j}$ are known (either from the calibration procedure or by diagonalisation of $M$ in a theoretical, numerical, or empirical way), there remain $N+1$ unknowns --- $m_o, \vecc{v}_o$ --- in \cref{eq: full_square}; they can be extracted using $ K \geq N+1$ measurements  $J_k = J(\vecc{v}_k), \ k=1,\ldots, K$ of the metric in $K$ test points $\vecc{v}_k$.
Thus, in the case of diagonal $\vecc{M}$, the test points $\vecc{c}_k, \ k=0,\ldots, N$ can be chosen as 
\begin{equation}
\vecc{c}_k = (0,\ldots, 1,\ldots, 0),
\end{equation}
where $1$ is present only on the $k$-th position.
The calculation of the optimal point $c_o$ from measurements $J_k$ is then straightforward.

\subsection{Example of aliasing in suboptimal basis}\label{seq: aliasing}

Consider the following  hypothetical example of wavefront correction with a deformable mirror able to reproduce any Zernike polynomial, and suppose we want to limit ourselves only to the correction of the first 6 polynomials\footnote{usually in microscopy, tip, tilt, and defocus (and of course piston) are discarded from the modal space as they correspond to the translation of the image in 3D space; here we kept them for the sake of simplicity---the example can be easily modified to include any other first $N'$ Zernike modes as the modal space and a higher order mode as the aberration} using the SM-based method, that is, using metric $J$ given by \cref{eq: sm discrete}.
Suppose also that we can calculate without error the second moment of PSF, \emph{i.e.} there is no noise, and the domain of $(x,y)$ (the camera size) is infinite.
Lukosz-Braat (LB) polynomials diagonalise the matrix given by \cref{eq: sm matrix} for a circular aperture for any $N$, so $m_{i,j} = \delta_{ij}, \ i,j = 1,\ldots,\infty$, where $\delta _{ij}$ is the Kronecker symbol; consequently they are often used as a theoretical basis for SM-based methods. 
The first 6 LB polynomials span
the modal space $F$   given by
\begin{equation}
F= \langle %
1, x, y, x^2, xy, y^2
\rangle,
\end{equation}
where $\langle \cdot \rangle$ stands for the linear span.
This is  the same space as spanned by the first 6 Zernike polynomials, 
so any aberration that belongs to this space would be corrected by the mirror and SM method with zero error.

In our example, let's take the incoming wavefront $\phi$ outside of space $F$, for instance a LB polynomial of a higher degree  $ B_4^0$ 
\begin{equation}\label{eq:B40}
\phi = B_4^0 =  6 r^4 - 8  r^2  +2.
\end{equation} 
As LB basis also diagonalises the metric $J$ for \emph{any} number of the first  terms, the value of metric $J$ will not change when applying any test aberration from the first 6 modes.
Thus, the second moment methods will not see the aberration and leave it without change.

\begin{figure}[t!]
	\centering
	\subfloat[PSF of the remaining aberration before (left) and after (right) correction by the second moments method 
	\label{fig:ex2:PSF}]{\includegraphics[width=0.485\linewidth]{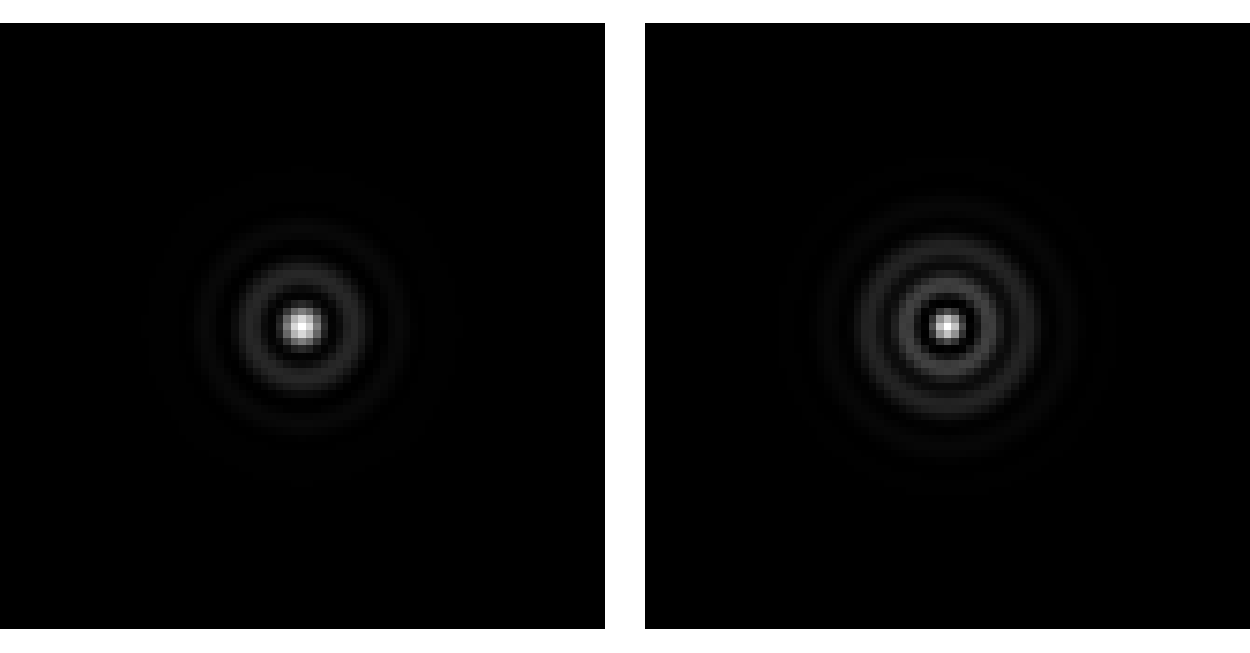}} \hfill
	\subfloat[Test image convolved with the PSFs from \protect\subref{fig:ex2:PSF}\label{fig:ex2:Image}]{\includegraphics[width=0.485\linewidth]{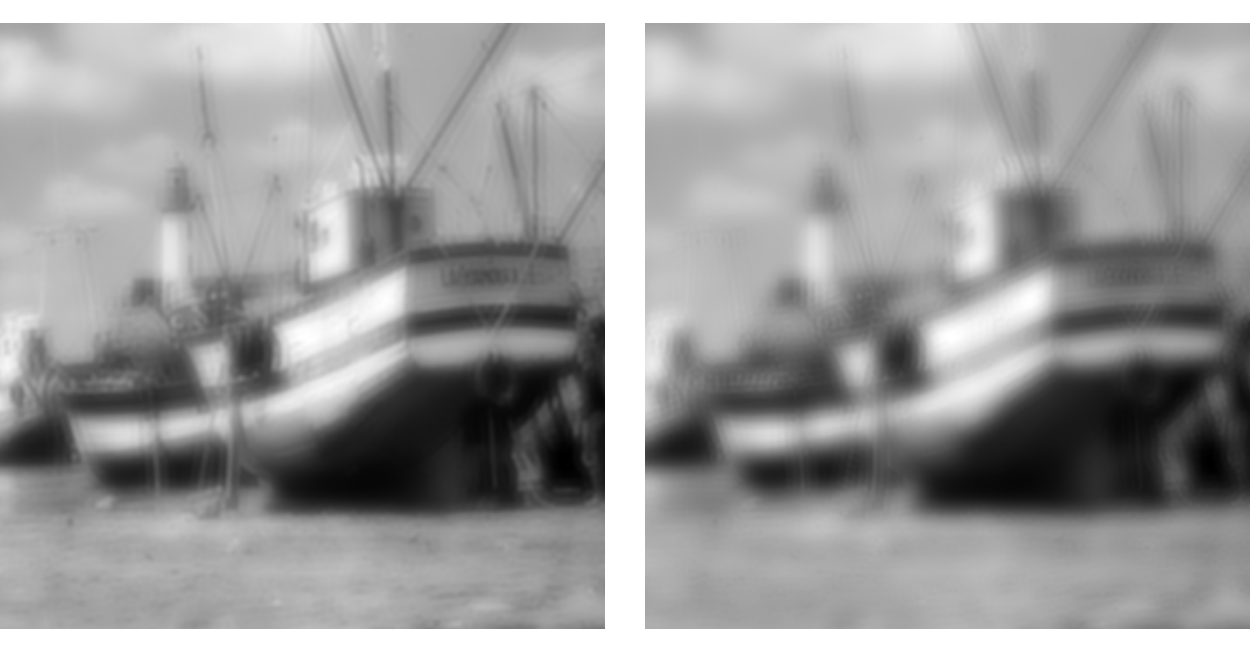}}\\
	\caption{Example of an aliasing error in the wavefront sensor-less method. Spherical aberration \cref{eq: z40} was used with an \emph{rms} amplitude of 1 rad. The second moment of the uncorrected PSF  (\protect\subref{fig:ex2:PSF}, left) is greater than the PSF corresponding to the \emph{rms} best-fit correction (\protect\subref{fig:ex2:PSF}, right), but the corrected image (\protect\subref{fig:ex2:Image}, right) contains more defocus. }
	\label{fig:example0}
\end{figure}

On the other hand, in the sense of \cref{eq: DM and not DM functions}, $\phi$ contains a non-zero part $\varphi_{DM} = -2 r^2 +1$ that should be corrected by the mirror. 
Thus, $B_4^0$ is not seen by the SM method and will be \emph{under-corrected}. 
Symmetrically,
the spherical aberration $Z_4^0$ from the space $\{\varphi_r\}$ of  uncorrectable terms of \cref{eq: DM and not DM functions} can be expressed as
\begin{equation}\label{eq: z40}
\phi' = 6 r^4 - 6 r^2 +1 = B_4^0 +(2r^2 -1),
\end{equation}
will be seen by the SM method as the defocus term $Z_2^0$ (see \cref{fig:example0}):
\begin{equation}
\phi \approx \varphi' = 2 r^2-1.
\end{equation} and will be \emph{over-corrected} by applying  the corresponding correction.

This example can be considered as an illustration of \emph{aliasing}, a manifestation of a ``high-order'' term $Z_4^0$ not belonging to the modal space $F$, as a non-zero ``low-order''  term $Z_2^0 \in F$, or equivalently, the presence of a non-zero correctable term in a aberration ``not-seen'' by the metric. 
In other words, the aliasing here was caused by the choice of space $F$ of the first $N'$ mirror responses, that was not orthogonal  to a high-order function from the basis that diagonalises the metric ($B_4^0$).

From the dot-product matrix  of the LB polynomials (see second line of Fig.~\ref{fig:modes}), it is clear that any subspace $F$ formed by a finite number of LB modes (not necessarily the first ones) will have an infinite number of non-orthogonal to it LB modes.
Moreover, the aliasing will remain for any other orthogonal basis in $F$ (e.g. a basis obtained by SVD of the first $N'$ modes with tip, tilt, and defocus excluded), as choice of basis does not change the space itself. Thus we can conclude that the nature of Zernike (or equivalently LB) polynomials does not allow the use of the decomposition of~\cref{eq: DM and not DM functions} without introducing an aliasing error. 

From our example, with the space $F$  formed by the first $N'$ modes of the LB basis, it is clear that the necessary condition for a basis diagonalising SM metric $J$ to be  aliasing-free   is that the functions from  its tail should be orthogonal to all the functions from its start. 
This can be generalised further to conclude that bases that diagonalise the metric but of which the dot-product matrix is not block-diagonal are not aliasing-free and are thus \emph{sub}-optimal. 
Therefore for use in indirect wavefront sensing methods, it would be more appropriate to call a basis ``optimal'' if it a) is  formed by a set of orthonormal functions and b) diagonalises the metric.
In this redefined optimal basis, decreasing the number of decomposition terms will only introduce a corresponding truncation error caused by the non-zero norm of $\varphi_r$; in a sub-optimal basis, an  additional aliasing error will also be present. 

The natural question is whether a defined in such a way optimal, aliasing-free, basis does exist for a given metric. 
The next section provides an example of a positive answer.

\subsection{Example of an aliasing-free basis}

It's relatively easy to find an example of an optimal basis for the second-moment metric and a \emph{square aperture}. This is provided by the trigonometrical basis, defined on a unit square $(x,y) \in [0,1] \times [0,1]$ as
\begin{equation}\label{eq:trig basis def}
f_{k l} (\vx) = \ee^{2 \pi \ii (k x + l y)}, \quad k, l \in \mathbb{Z},
\end{equation}
or its real-valued equivalent $\{\sin (k x + l y), \cos( k x + l y)\}$.
The basis is orthogonal  on the unit square, and it is easy to check that it diagonalises the matrix given by \cref{eq: sm matrix} as its gradients are also orthogonal on the unit square, so the truncated Fourier series of the decomposition of $\phi$ in this basis 
can be used for alias-free indirect wavefront sensing.

\section{Derivation of an aliasing-free basis for any aperture}\label{sec: necessary}

Now we're going to find an optimal basis for the SM metric for  a circular aperture, for which
both  the modes  and their gradients should form two orthogonal sets on a unit disk.
Moreover, we find a sufficient and necessary condition for a basis to be optimal
for second-moment based methods 
on \emph{any} aperture.
For this, consider the following mathematical formalisation of the problem.

Consider the Hilbert space $L = L_2[P]$, where $P$ is the phase domain, that is, the pupil of the system, with the ``standard'' $L_2$ inner product defined as
\begin{equation}\label{eq: standard inner product}
(f,g) = \int_P f(\vx) g(\vx) \dd \vx,
\end{equation}
and let  $F \subset L$ be a subspace, spanned by some $N'$ functions or modes $f_n(\vx), \ n=1,\ldots, N'$ 
\begin{equation}\label{eq:modal space}
F = \langle f_1, \ldots, f_{N'} \rangle.
\end{equation}
Let $\phi \in L$ be some function\footnote{for the sake of brevity, from now on the dependence on $\vx$ will be omitted if it is not causing any problems, and tensor notation will be used, so a variable with an upper letter index $c^n$ denotes a column vector $c^n = [c^1, \ldots, c^{N'}]^T$, and a variable with a lower index denotes a row vector; a repeating upper and low index denotes summation}
that is approximated by an  element $\varphi$ of $F$:
\begin{equation}\label{eq: modal representation}
\phi(\vx) \approx  c^0 + \varphi(\vx) = c^0 + \sum_{n=1}^{N'} c^n f_n (\vx) \equiv c^0 + c^n f_n,
\end{equation}
where the piston term $c^o$ is usually separated as it has no influence on the PSF.

With $\phi$ representing the incoming wavefront, and $f_1,\ldots,f_{N'}$ representing the decomposition modes, the 
modal wavefront reconstruction 
may now be represented as an optimisation problem in the space $L_2[P]$  
\begin{equation}\label{eq:optimisation} 
\min_{c^0, c^n} \norm{\phi - c^0 - c^n f_n }_2^2.
\end{equation}

Without loss of generality, we can consider the set  $f_n$ to be the first $N'$ functions of a full and orthonormal basis $f_n,\ n=1,\ldots, \infty,$
\begin{equation}
(f_n, f_{n'}) = \delta_{n,n'},
\end{equation} 
where  $\delta$ is Kronecker's symbol.
Then the best approximation $\varphi$  of $\phi$ is given by truncation of its  decomposition by the basis functions
\begin{equation*}
\phi = \sum_{n=1}^{\infty} c^n f_n \; \Rightarrow \varphi = \sum_{n=1}^{N'} c^n f_n,
\end{equation*}
where the decomposition coefficients $c^n$ are obtained by the inner product of $\varphi$ with the basis functions:
\begin{equation}\label{eq: Fourier coefficient}
c^n = (f^n,\phi),
\end{equation}
and are known as the (generalised) Fourier coefficients. 
From the orthonormality of the basis, we get the Euclidean norm of the approximation error $\varepsilon = \phi - \varphi$ as
\begin{equation}
\norm{\varepsilon}^2= \norm{\phi}^2 -\norm{\varphi}^2 = \sum_{n=N'+1}^{\infty} (c_n)^2 = \norm{\phi}^2 -  \sum_{n=1}^{N'} (c_n)^2  .
\end{equation}

From \cref{eq: SM integral on v} it follows that the second-moment-based methods approximate the gradient of the incoming wavefront with the gradients of the basis function, solving thus another optimisation problem
\begin{equation}\label{eq:optimisation G} 
\min_{ c^n} \norm{\nabla\phi  - c^n \nabla f_n }_2^2.
\end{equation}
Here, the piston term $c_0$ is cancelled by the gradient operator $\nabla$.

Disregarding the piston term, the optimal (alias-free) basis should provide the same solutions for the problems \cref{eq:optimisation} and \cref{eq:optimisation G}.

Let us remove the piston term from both problems by redefining $L$ as a factor space by the gradient kernel: $L=L_2[P]/\{\phi:\phi=\textrm{const}\}$.
In the factor space $L$, the optimisation problem \cref{eq:optimisation} is equivalent to the decomposition of  $L$  into the direct sum of $F$ and its orthogonal complementation $O= F^{\perp}$:
\begin{equation}\label{eq: modal and error space}
L = F \oplus O, 
\end{equation} 
where  $\varphi \in F$ and $\varepsilon \in O$ are the orthogonal projections of $\phi$ on the modal space and its orthogonal complementation, and
\begin{equation}
\norm{\phi}^2 = \norm{\varphi}^2 + \norm{\varepsilon}^2.
\end{equation}

Let us also introduce another inner product in $L$, defined as the ``standard'' inner product\footnote{the inner product $(\nabla f, \nabla g)$ is defined here for the vector-functions from $L\times L$ in a natural way: $(\nabla f, \nabla g) =  (\frac{\partial f}{\partial x}, \frac{\partial g}{\partial x}) + (\frac{\partial f}{\partial y}, \frac{\partial g}{\partial y}) $}
of the \emph{gradients}:
\begin{equation}\label{eq: gradient inner product}
(f,g)_\nabla \stackrel{\textrm{def}}{=}  (\nabla f, \nabla g) =
\int_P \nabla f(\vx)\cdot \nabla g(\vx) \dd \vx,
\end{equation}
where $\vecc a \cdot \vecc b$ denotes the ``standard'' inner product in $\mathbb R^2.$ One can easily verify that in the factor space $L$, relation \eqref{eq: gradient inner product} satisfies all properties of an inner product,  which we will call the ``gradient-dot product'', and therefore we can define a corresponding ``gradient-norm'' as
\begin{equation}
\norm{f}_\nabla = \sqrt{(f,f)_\nabla} \equiv \sqrt{\int_P \abs{\nabla f}^2 \dd \vx}.
\end{equation}

With the gradient-norm, the minimisation problem of \cref{eq:optimisation G} can be written as 
\begin{equation}\label{eq:optimisation G norm} 
\min_{c^n} \norm{\phi - c^n f_n}_\nabla,
\end{equation}
and solving this problem is equivalent to finding the orthogonal complementation of $F$  with respect to the gradient-dot product: 
\begin{equation}\label{eq: modal and error space G}
L = F \oplus_\nabla O_\nabla. 
\end{equation}

\begin{figure}[thb]
	\labellist
	\small
	\hair 2pt
	\pinlabel {$L$} [ ] at 22 135
	\pinlabel {$F$} [ ] at 38 67
	\pinlabel {$O$} [ ] at 111 144
	\pinlabel {$O_\nabla$} [ ] at 164 142
	\pinlabel {$\phi$} [ ] at 182 110
	\pinlabel {$\varphi$} [ ] at 182 66
	\pinlabel {$\varphi'$} [ ] at 160 77
	\pinlabel {$\varepsilon'$} [ ] at 135 102
	\pinlabel {$\varepsilon$} [ ] at 110 111
	\endlabellist
	\centering
	\includegraphics[width=0.61\linewidth]{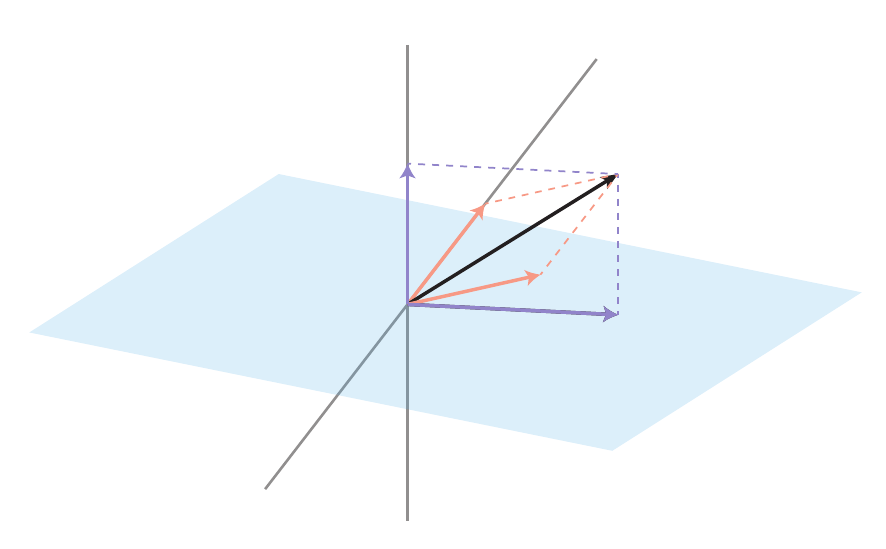}
	\caption{Modal space $F$ and its two orthogonal complementations $O$ and $O_\nabla$  in two different inner products. $\varphi$ and $\varphi '$ are the best approximations of $\phi$ by the modal functions in the sense of the optimisation problems \eqref{eq:optimisation} and \eqref{eq:optimisation G}, respectively. They are given by the orthogonal projections $\Pr_F (\phi),\Pr_{F,\nabla} (\phi) $ in two different inner products, and  they are equal if and only if $O= O_\nabla.$ }
	\label{fig:spaces}
\end{figure}
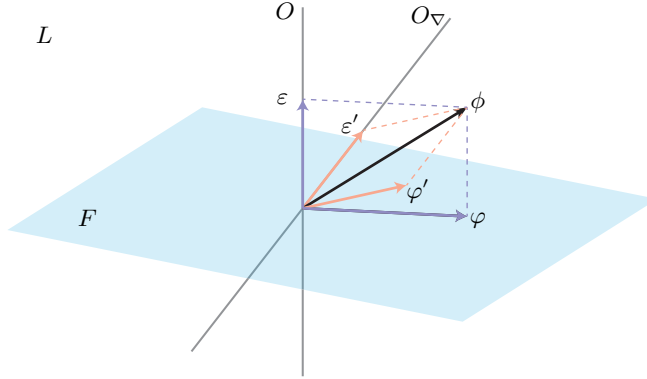

Obviously, as \cref{eq:optimisation} and \cref{eq:optimisation G norm} use different norms to minimise the residue, their solutions do not necessarily  match. 
It is easy to prove (see Fig.~\ref{fig:spaces}) that they would be the same if and only if the two orthogonal complementations of $F$ are the same, that is \begin{equation}\label{eq: alias-free condition}
O=O_\nabla .
\end{equation}

Indeed, if for any $\phi$ its projections on $F$ in both inner products match, \emph{i.e.} $\varphi  = \varphi '$, then consequently its approximation errors match too, \emph{i.e.} $\varepsilon = \varepsilon'$. Then for any vector $x\in O_\nabla$, considering its approximation errors, one has  $ x= \varepsilon ' = \varepsilon \in O$, and thus $O_\nabla \subseteq O$. In the same way,  $O \subseteq O_\nabla$ and thus $O_\nabla = O.$

Vice versa, suppose that $O_\nabla = O.$ Then from Eqs.(\ref{eq: modal and error space}) and (\ref{eq: modal and error space G}) and from the definition of the direct sum, we get that $\varphi = \varphi '$.

This explains why optimisation of the Zernike basis by changing to LB or SVD modes may introduce an aliasing error.
By performing this optimisation, we re-orthogonalise the basis modes in $F$ with respect to the dot-gradient product and thus also implicitly introduce a new inner product, which changes the orthogonal complementation from $O $ to $O_\nabla$.
For instance, no first $N$ terms of the Zernike modes considered in the example above can form an optimal basis for the second-moment-based moments. 
Noll \cite{Noll76}  has calculated the coefficients of decomposition of the $x$- and $y$-components of the gradient of Zernike polynomials through themselves, and
from his calculation it follows that
\begin{equation}\label{eq: Zernike grad orth}
(Z_n^0,Z_{n+2}^0)_\nabla \neq 0 \ \forall n, \ n \text{ is even}.
\end{equation}
Hence, if $n$ is the maximal order of spherical aberration from $F$, then $Z_{n+2}^0 \notin O_\nabla$. But as $Z_{n+2}^0 \in O$, this means $O \neq O_\nabla$.

Let us call the basis  $f_n, n=1,\ldots,\infty$ in $L$ ``optimal'' if its elements are both orthogonal and gradient-orthogonal:
\begin{equation}
(f_i, f_j) = (f_i, f_j)_\nabla = 0 \  \text{if} \ i\neq j.
\end{equation}
For the modal space $F$ formed by any $N'$ functions of the basis $F= \langle f_{i_1},\ldots,f_{i_{N'}} \rangle $, condition~\eqref{eq: alias-free condition} is obviously satisfied, and thus limiting the number of decomposition modes of the optimal basis will not introduce an aliasing error.
In this basis, the solution to the optimisation problem~\eqref{eq:optimisation G norm} is given in a way similar to \cref{eq: Fourier coefficient}:
\begin{equation}\label{eq: Fourier coefficient G}
c'_n =\frac{ (f_n,\phi)_\nabla}{(f_n,f_n)_\nabla}.
\end{equation}

Now we are going to find a necessary condition for a basis to be optimal  
for an arbitrary  aperture.
Suppose $F$ has matching orthogonal complements $O$ and $O'$. Then for any vector $\phi$, its orthogonal projection on $F$ should match in both inner products
\begin{equation}
\Pr_F (\phi) = \Pr_{F,\nabla} (\phi).
\end{equation}
From this it follows that the $L_2$-scalar product and the gradient-dot product of any test function $\phi$ and any modal vector $f_n$ should be proportional to each other, or more accurately, if $k_n$ is the gradient-norm of a basis vector $f_n$, then
\begin{equation}
(\phi, f_n) = \frac{1}{k_n^2} (\phi, f_n)_\nabla.
\end{equation}
Let us take a Dirac $\delta$  as a test function, $\phi(\vx) = \delta(\vx -\vx_0)$ for some $\vx_0 \in P$. 
By definition of the gradient-dot product ~\cref{eq: gradient inner product} and by Stokes theorem, we get
\begin{multline}
f_n(\vx_0) = ( \delta(\vx -\vx_0), f_n) = \frac{1}{k_n^2} ( \delta(\vx -\vx_0), f_n)_\nabla \\ 
=  \frac{1}{k_n^2} ( \nabla \delta(\vx -\vx_0), \nabla f_n) = - \frac{1}{k_n^2} (  \delta(\vx -\vx_0), \nabla^2 f_n) \\ 
= -\frac{1}{k_n^2} \Delta f_n (\vx_0) \quad \forall \vx_0 \in P,
\end{multline}
that is to say, $f_n$ should be the eigenfunction of the Laplace operator, and is given by the solution of the Helmholtz equation in the aperture $P$:
\begin{equation}\label{eq: Helmholz equation}
\Delta f_n + k_n^2 f_n = 0
\end{equation}
for some $k_n \in \Real.$
This equations comprises the necessary condition for $\{f_n\}$ to form an optimal basis.

As the eigenfunctions of the Laplace operator satisfying to the homogeneous boundary condition form an orthogonal basis, we have just proved that \emph{any} optimal modal space for the gradient operator is given by the solutions to \cref{eq: Helmholz equation} with some homogeneous boundary conditions.
The trigonometric basis for the square aperture of~\cref{eq:trig basis def} is a particular case of such a  solution
to \cref{eq: Helmholz equation}, with periodic boundary conditions.
For a circular aperture, this is a classical problem of mathematical physics, with the analytical solution  bounded at $\vx =0$ phase given in the polar coordinates $\vx = (r,\theta)$ by
\begin{equation}\label{eq: Helmholtz circular solution}
	f_{m,n}(r, \theta) = J_m(k_{m,n} r) 
	 \genfrac{}{}{0pt}{1}{\cos}{\sin} (m \theta),
\end{equation}
where $J_m(r)$ are the Bessel functions of the first kind, and $k_{m,n}$ some constants depending on the boundary conditions.
It's interesting to note that this set of functions, referred as the Bessel circular functions or the membrane vibration modes, has already popped up recently in the literature\cite{Trevino2013, Hadipour2016} as being more suitable than  Zernike polynomials for some applications, but, to our knowledge, was not yet considered as optimal for wavefront sensor-less methods.


\section{Numerical simulation}\label{sec:simulations}


\begin{figure*}[tbhp!]
	\centering
	\begin{tabular}{@{}lc@{}}
		\toprule
		\rotatebox[origin=c]{90}{Zernike} & \raisebox{-.5\height}{\includegraphics[width=.9\linewidth]{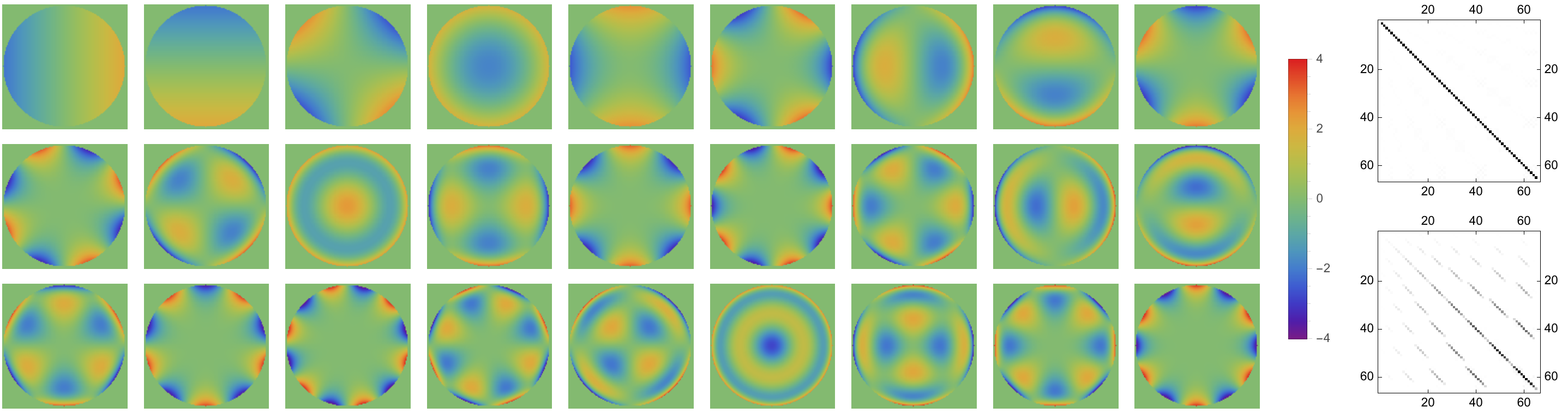}} \\[.61\baselineskip] \midrule
		\rotatebox[origin=c]{90}{Lukosz-Braat} & \raisebox{-.5\height}{\includegraphics[width=.9\linewidth]{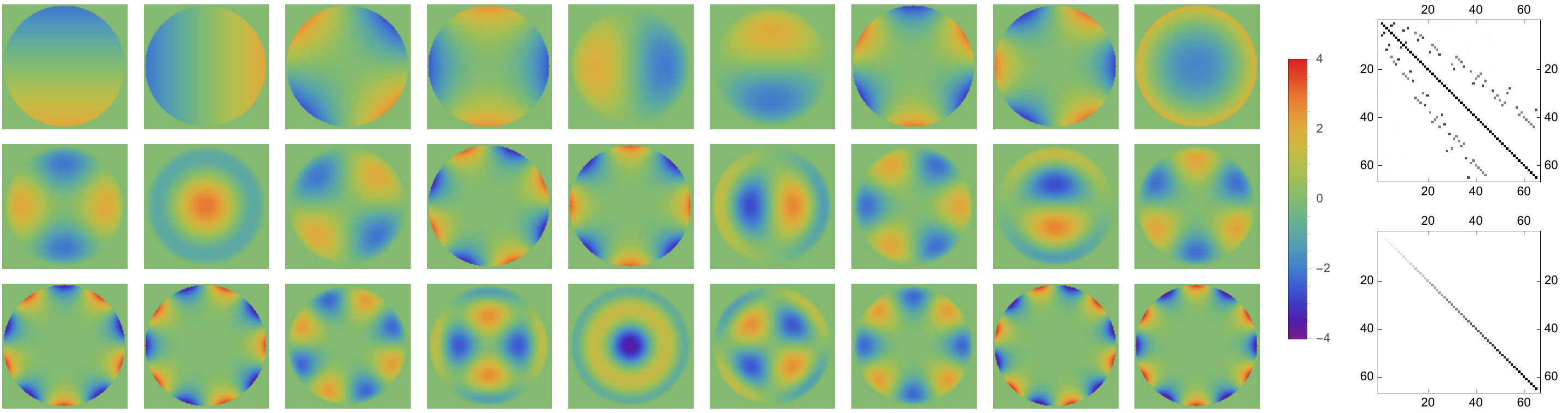}} \\[.61\baselineskip] 
		\midrule
		\rotatebox[origin=c]{90}{\parbox{2.2cm}{SVD modes, \\first 4 orders}} & \raisebox{-.5\height}{\includegraphics[width=.9\linewidth]{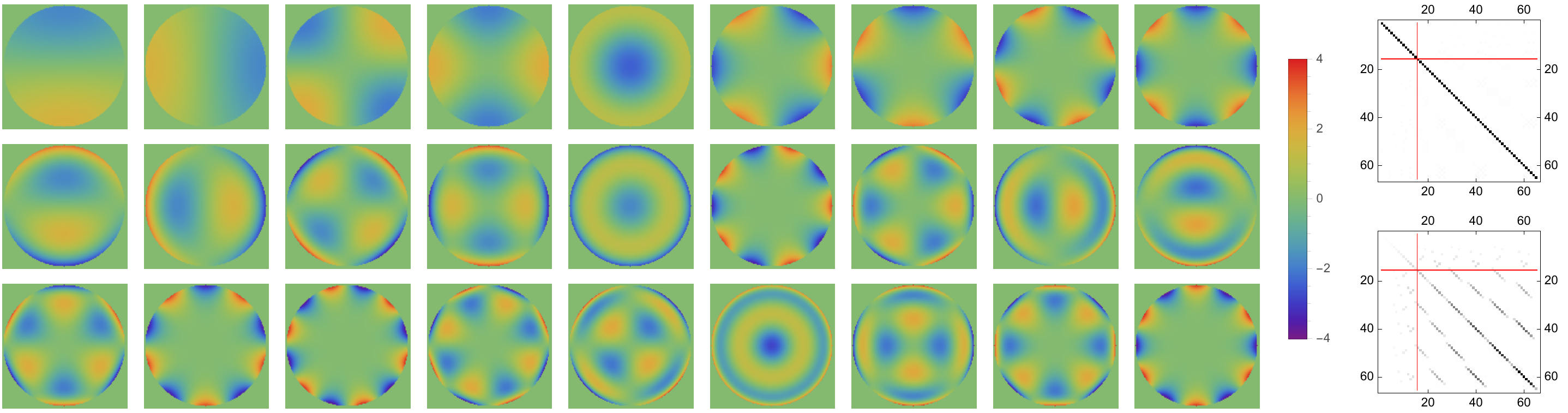}} \\[.61\baselineskip] 
		\midrule
		\rotatebox[origin=c]{90}{\parbox{2.2cm}{SVD modes, \\first 10 orders}} & \raisebox{-.5\height}{\includegraphics[width=.9\linewidth]{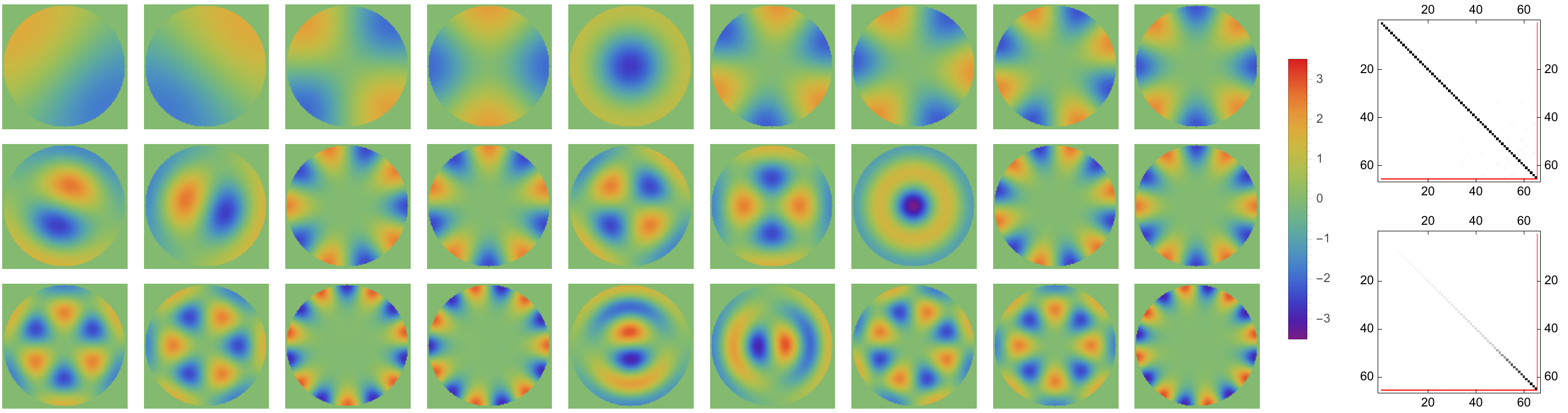}} \\[.61\baselineskip] 
		\midrule
		\rotatebox[origin=c]{90}{\parbox{2.5cm}{Optimal modes,\\ Neumann}} & \raisebox{-.5\height}{\includegraphics[width=.9\linewidth]{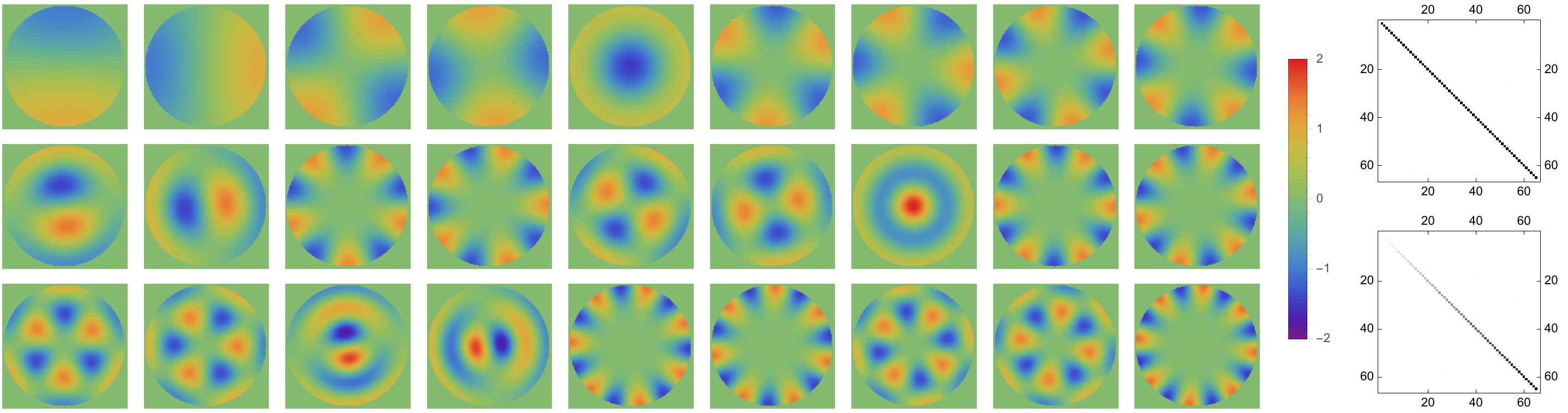}}
		\\[.61\baselineskip] 
		\bottomrule
	\end{tabular}
	\caption{First 27 terms of the modes used for the approximation, top to bottom: Zernike polynomials, Lukosz-Braat polynomials, SVD modes obtained from the first 4 orders of the Zernike polynomials, SVD modes obtained from the first 10 orders of Zernike polynomials, optimal modes satisfying the  Neumann
		boundary conditions. On the right of each set, the non-zero elements of the matrices of dot product $\vecc M$ (top) and gradient dot products $\vecc{ M'}$ of the first 65 terms are shown.
	The red lines in the matrix plots for SVD modes indicate the subspace on which SVD has been performed.}
	\label{fig:modes}
\end{figure*}

In this section we compare the accuracy of aberration correction using sub-optimal and optimal bases  with a numerical simulation.

It is not evident whether a basis satisfying the optimality criteria would suit a particular application, because optimality as defined in this paper means only that the result of two optimisation problems would be the same and no aliasing will be introduced by using the gradient norm. 
In other words, the basis might be optimal for a given wavefront sensing method, but not for representing a typical aberration of an application.
The magnitude of the aliasing error is also defined by the statistical properties of the aberration and  by the speed of convergence of the generalised Fourier series. 
For instance, the eigenfunctions defined with the Dirichlet boundary conditions seem \emph{a priori} to be less suitable for  aberrations with non-constant values on the boundary.

As test cases, we have used a) wavefront aberrations produced by mouse oocyte cell as an example of specimen induced aberration in microscopy~\cite{Schwertner2004-charachterisation} and b) wavefront aberrations produced by Kolmogorov's turbulence, covering two important applications of AO\footnote{see Ref.~\citenum{Trevino2013} for the fitness analysis for ophthalmic applications}.
Although the iterative nature of the indirect wavefront sensing methods makes them less suitable for the  case of the atmospheric turbulence, we have included this example for comparison purposes and to get an impression of the extent to which the improvement given by the optimal modes depends on the wavefront statistics.

\begin{figure}[thbh!]
	\centering
	\includegraphics[width=1\linewidth]{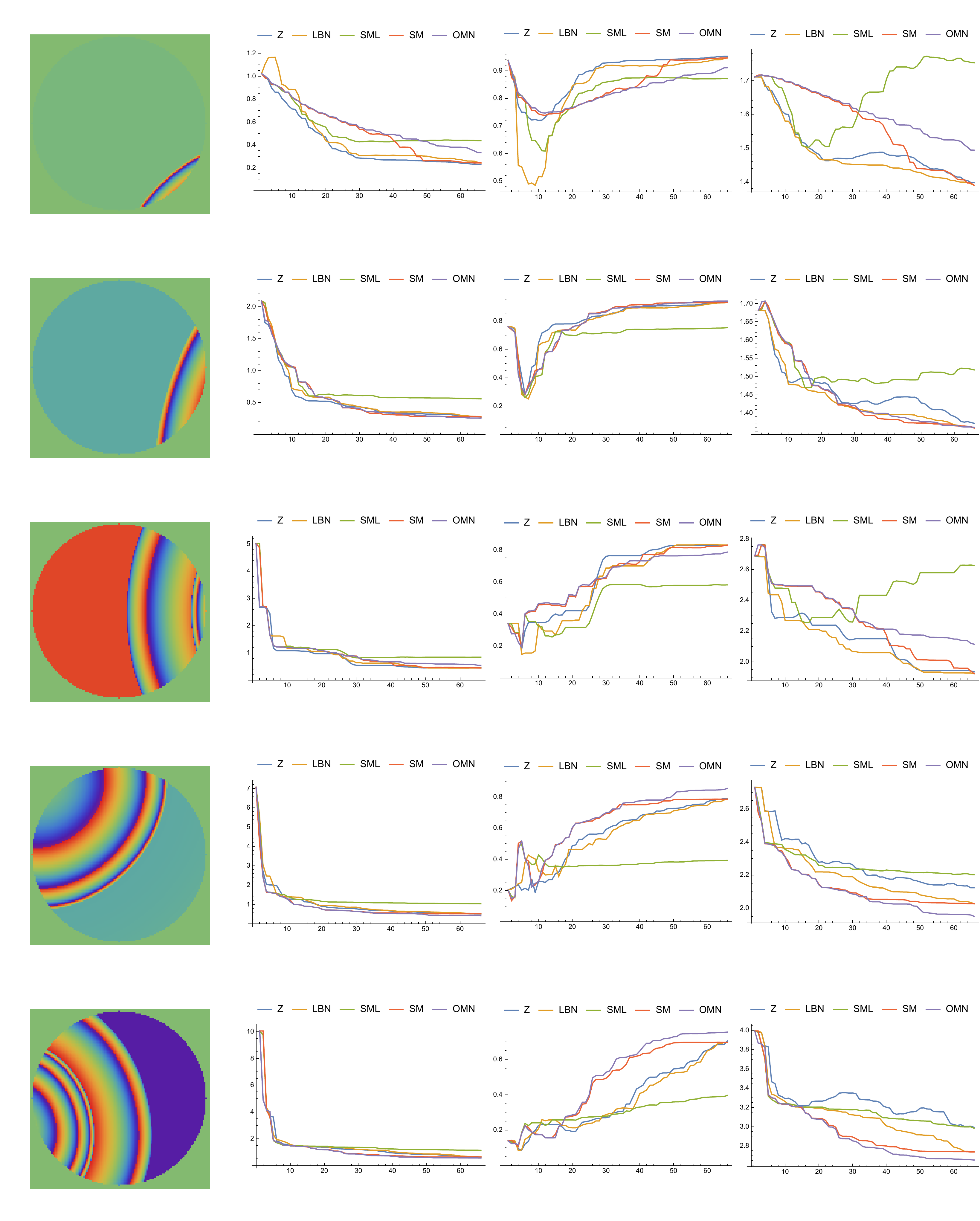}
	\caption{5 oocyte-induced aberrations of growing strength (left column) and the quality of their approximation  with different bases shown as plots of (left to right): a) rms b) Strehl ratio c) normalised second moment of the remaining aberration vs the number of used modes $N'$.
	The aberrations are shown as wrapped phase using the following colour scheme: \raisebox{-1\baselineskip}{\includegraphics[height=2\baselineskip]{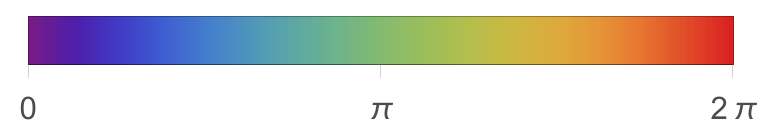}}.}
	\label{fig:oocytequality}
\end{figure}

\begin{figure}[thbp!]
	\centering
	\includegraphics[width=1\linewidth]{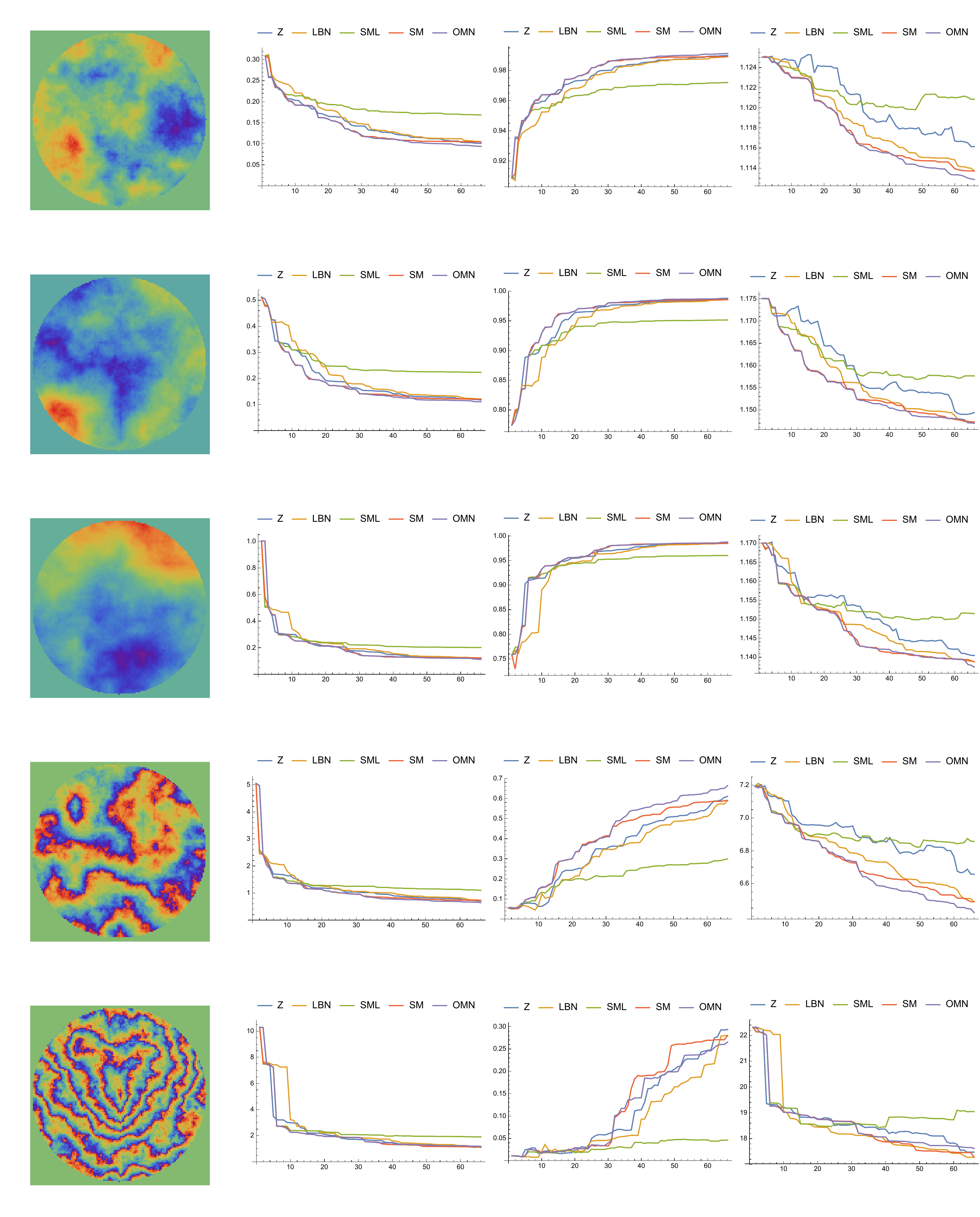}
	\caption{5 turbulence-induced aberrations of growing strength (left column) and the quality of their approximation  with different bases shown as plots of (left to right): a) rms b) Strehl ratio c) normalised second moment of the remaining aberration vs  the number of used modes $N'$.
		The aberrations are shown as wrapped phase using the following colour scheme: \raisebox{-1\baselineskip}{\includegraphics[height=2\baselineskip]{colorbar3}}.}
	\label{fig:turbquality}
\end{figure}

The oocyte phase screens were obtained by the procedure described in Ref.~\citenum{Schwertner2004} with the same parameters as its authors used for the results demonstrated in their Figure 5. 
The Kolmogorov phase screens were calculated by the method described in Ref.~\citenum{tub_sim} with 20 phase screens for the ratios $D/r_0 = 25/ i$  with integer $i$ in the range from 1 to 25. 
The typical phase screens for both test cases are shown in the left columns of 
\cref{fig:oocytequality,fig:turbquality}.
The screens are shown wrapped for the presentation purposes only; unwrapped phase screens were used for approximations.
500 random phase screens for each of the cases were generated; several aberration-free phase screens were generated 
by the method of Ref.~\citenum{Schwertner2004}; they were discarded, resulting in a total of 459 phase screens for the oocyte-case.

For each of the phase screens,
we compared the numerical results of its approximation 
by the gradients of the first 10 orders (= 65 terms) of the 
a) Zernike polynomials (denoted in the plots as Z);
b) LB polynomials  (LB);
c) SVD modes obtained from the first 4 orders of the Zernike modes with respect to the gradient-dot product (SML);
d) SVD modes obtained from  all 10 orders of the Zernike modes with respect to the gradient-dot product (SM), and eigenmodes of the Laplacian satisfying  
either to
d) Neumann boundary conditions (OMN) or
e) Dirichlet boundary conditions (OMD)\footnote{the results for the OMD case are not present in this paper, but are available in the supplementary material~\cite{oleg_soloviev_2017_801515}}.

Each of the basis functions and aberrations was discreetized on a $128 \times 128$ grid, with a circular aperture of a 125 pixels diameter (see \cref{fig:modes}). 
The Zernike polynomials were calculated using their analytical formulae and normalised with respect to their $\ell_2$ norm on the aperture. 
Their gradients were approximated by first-order finite differences inside the aperture; the values on the aperture boundary were zeroed. 
The Lukosz-Braat polynomials were calculated numerically through Cholesky decomposition of the gradient-dot product matrix $\vecc{M'_Z}$ of the Zernike modes (because  the gradients are not defined on the boundary, this provided more accurate results for the chosen resolution than calculation  using analytical formulae). 
The SVD modes were obtained by SVD decomposition of $\vecc{M'_Z}$. 
The optimal modes were obtained in \emph{Mathematica} software with the \texttt{NDEigensystem} function with the corresponding boundary condition.  
All modes were, if necessary, first renormalised with respect to the aperture and then reordered by increasing their gradient-norm. 

The dot product and the gradient dot product matrices $M$ and $M'$ were calculated numerically for the discreetized bases for illustration purposes (see right column of \cref{fig:modes}); the matrices allow to see the key difference between seemingly similar basis function shapes. 
Please note how SVD modes diagonalise both matrices \emph{only} for the orders from which they were calculated (the top left minor formed by the red lines in the matrix plots). 

For each of the phase screens and each of the bases, we have calculated
\begin{inparaenum}[a)]
	\item its \emph{rms} error by approximating it by the first $N'$ Zernike terms, 
	\item the \emph{rms} error of its approximation by the gradients of the first $N'$ terms of the basis,
	\item the Strehl ratio\footnotemark, and
	\item the second moment of the PSF of the remaining aberration (normalised to that of the diffraction-limited PSF).
\end{inparaenum}
\footnotetext{
	The Strehl ratio was defined as the maximum intensity of the central pixel, normalised to the diffraction-limited maximum. 
	This definition results in a specific dip of the plots for the oocyte-produced small aberrations, like in the top rows of \cref{fig:oocytequality}: for instance, correction of tilt only would move the brightest pixel further away  from the image centre. 
} 

The results for 5 representative aberrations of various strengths for each set are shown in Figs.~\ref{fig:oocytequality} and~\ref{fig:turbquality}.

\begin{figure*}[tb!]
	\centering
	\includegraphics[width=1\linewidth]{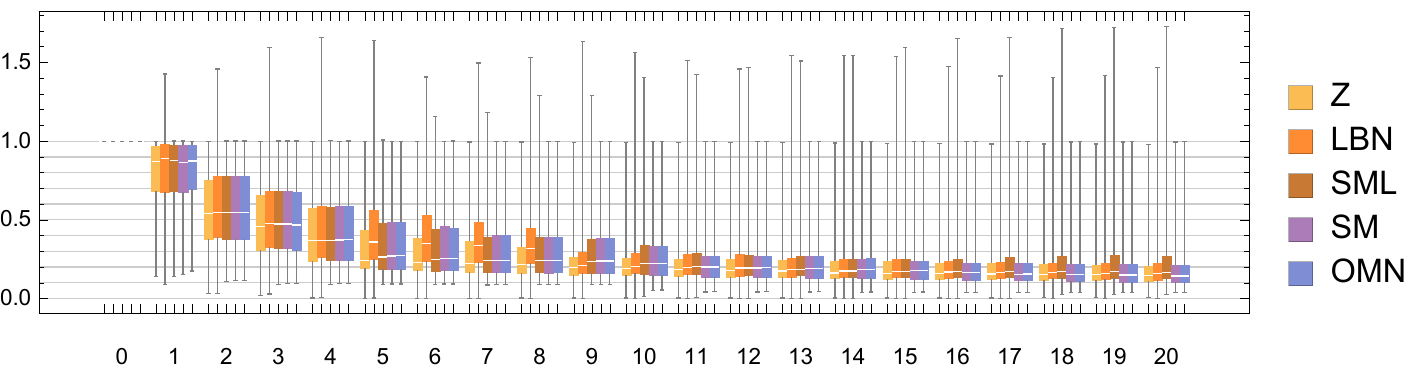}\\
	\includegraphics[width=1\linewidth]{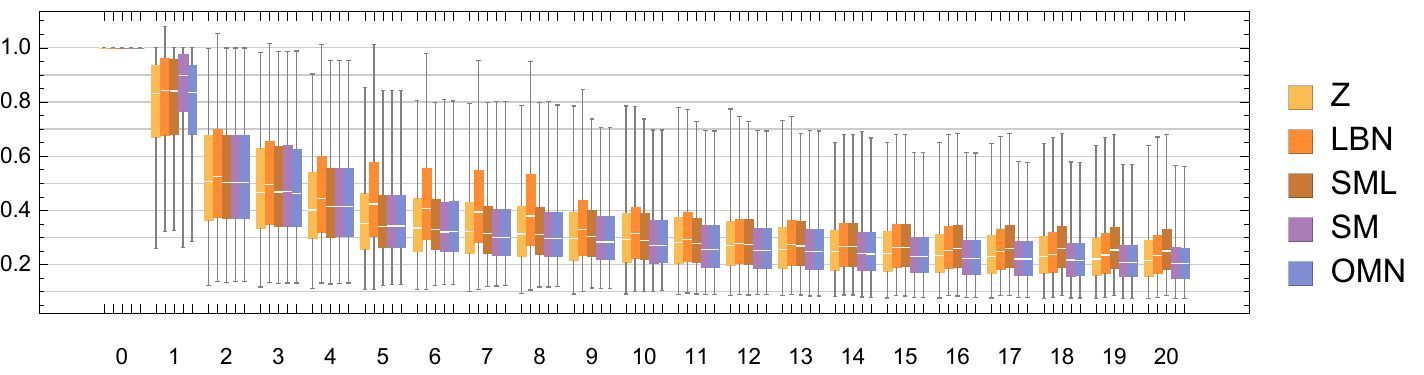}
	\caption{Relative \emph{rms} of the approximation error
		using the first $N', N'\leq 20$ modes  of the Zernike (Z) and Lukosz-Braat (LBN) polynomials, SVD modes obtained with respect to the gradient-dot product from the first 4 (SML) and the first 10 (SM) orders of Zernike polynomials, and the optimal modes with the Neumann boundary conditions (OMN). 
		The top plot is for the oocyte-induced aberrations, the bottom plot is for the turbulence phase screens.
		The boxes show the median value and quartiles.
		The phase screens and the Mathematica code used for the calculations are available at Zenodo~\cite{oleg_soloviev_2017_801515}.}
	\label{fig:results}
\end{figure*}

Figure~\ref{fig:results} presents the box-plot for the relative \emph{rms} error for the ensembles of the phase screens; it is shown only for the first third of the modes for better visibility.

From the results shown in Fig.~\ref{fig:results}, it is clear that the  optimal modes obtained from solution of Helmholtz equation with Neumann boundary conditions (denoted as OMN) provide on average better results for smaller dimension of the modal space (first 4 orders of Zernike, denoted as SML), and
similar approximation quality to that of the numerically calculated SVD modes obtained from the first 10 orders of Zernike polynomials (SM).
This result can be explained by the fact that for large enough $N$, the part of the aberration not covered by the span of the basis function is small, and correspondingly the aliasing error introduced by it is small as well.

\section{Discussion and conclusions}

While the results of \cref{sec: necessary} describe the best analytical basis for SM-based methods, this basis can be difficult to implement in practice because of alignment errors, necessity of exact calibration of the adaptive element and so on, so SVD modes are used commonly.
Numerical results of \cref{sec:simulations} show that
singular modes obtained from the large number of Zernike polynomials look similar to the optimal modes (\cref{fig:modes}), which can be seen as the limiting case of the SVD-ing with the number of modes going to infinity. 
This is also confirmed by the similarity of the results (see \cref{fig:results}, cases SM and OMN).

The optimal basis for a given aperture is not unique (for instance, both OMN and OMD are optimal), 
and thus the question of a \emph{proper} optimal basis \emph{for a given application} remains open. 
Similarity of SVD and optimal bases suggests that an approximation of the optimal basis can be used. 
For instance, for a piecewise smooth aberration like that of the oocyte-induced aberrations, SVD orthogonalisation of Gaussian radial base functions might give better results.
Moreover, as one can see by the first row of \cref{fig:oocytequality}, for a sparse aberration present only at one edge of the aperture and a small number of correction modes, the Lukosz-Braat functions give better results.
This can be explained by the specific shape of the aberration, which has 70 per cent of its power uniformly contributed by the first 30 Zernike polynomials.

An important conclusion on the practical use of SM-based wavefront sensorless methods can be drawn from   \cref{sec: necessary}.
To minimise the aliasing effect, the response functions of  the adaptive element should be a good approximation of some optimal basis, \emph{i.e.} approximately satisfy \cref{eq: Helmholz equation}.
Since the response functions of a membrane deformable mirror, a bi-morph mirror, and a free-edge mirror with push-pull actuators  satisfy either Poisson or biharmonic equations with a piece-wise constant right-hand part~\cite{Vdovin2008}, none of them, nor any basis obtained by orthogonalisation of their gradients, can form a true optimal basis, and thus the use of low-order deformable mirrors for correction with a second-moment based method will suffer from aliasing effects. 
Again, some approximation to the optimal basis can still be obtained with these types of mirrors with a large number of actuators, \emph{e.g.} a photo-controlled deformable membrane mirror~\cite{Bonora2012}.
For a low-order adaptive element, the layout of the actuators should be optimised in order to get response functions closer to the optimal modes (and not to Zernike modes, for instance).



To conclude, in this paper
we have derived the necessary condition for an analytically defined basis to be optimal, that is, aliasing-free, for wavefront sensorless methods based on the second moment of intensity, namely that each function of the basis should satisfy the Helmholtz equation, and we have provided examples of such a basis for square and circular apertures. 
The analytical optimal basis can be considered as a limiting case of a numerically computed optimal basis via SVD of the gradient-dot product when the number of basis functions is sufficiently large. 
The proposed analytical modes can be useful when controlling an adaptive element with a large number of actuators, as probe modes (starting point) for building an empirical optimal basis, and when designing low-order adaptive elements for SM-based methods.


\section*{Acknowledgements}

The author is grateful to G.~Vdovin, M.~Verhaegen, and D.~Wilding (TU Delft) for their attention to the work and valuable comments and to D.~Soloviev for proofreading the manuscript. 

\bibliographystyle{plain}
\bibliography{optimalBasis} 

\begin{thebibliography}{10}

\bibitem{Bonora2012}
S.~Bonora, D.~Coburn, U.~Bortolozzo, C.~Dainty, and S.~Residori.
\newblock {High resolution wavefront correction with photocontrolled deformable
  mirror}.
\newblock {\em Opt. Express}, 20(5):5178--88, 2012.

\bibitem{Booth2007}
Martin~J. Booth.
\newblock {Wavefront sensorless adaptive optics for large aberrations}.
\newblock {\em Opt. Lett.}, 32(1):5, 2007.

\bibitem{Booth2014}
Martin~J. Booth.
\newblock {Adaptive optical microscopy: the ongoing quest for a perfect image}.
\newblock {\em Light: Science {\&} Applications}, 3(4):e165, 2014.

\bibitem{Braat1987}
Joseph Braat.
\newblock {Polynomial expansion of severely aberrated wave fronts}.
\newblock {\em J. Opt. Soc. Am. A}, 4(4):643, 1987.

\bibitem{Debarre:07}
Delphine D\'{e}barre, Martin~J. Booth, and Tony Wilson.
\newblock Image based adaptive optics through optimisation of low spatial
  frequencies.
\newblock {\em Opt. Express}, 15(13):8176--8190, 2007.

\bibitem{Debarre2009}
Delphine D\'{e}barre, Biru Wang, Tony Wilson, and Martin~J. Booth.
\newblock {Optimum schemes for wavefront sensorless adaptive optics in
  microscopy}.
\newblock {\em Proc. SPIE}, 7209, 2009.

\bibitem{Facomprez2012}
Aur{\'{e}}lie Facomprez, Emmanuel Beaurepaire, and Delphine D{\'{e}}barre.
\newblock {Accuracy of correction in modal sensorless adaptive optics}.
\newblock {\em Opt. Express}, 20(3):2598, 2012.

\bibitem{Hadipour2016}
Mousa Hadipour, Murat Tahtali, and Andrew~J. Lambert.
\newblock Vibrating membrane mirror concept for adaptive optics.
\newblock {\em Proc. SPIE}, 9912:99123O--99123O--9, 2016.

\bibitem{tub_sim}
R.G. Lane, A.~Glindemann, and J.C. Dainty.
\newblock Simulation of a {K}olmogorov phase screen.
\newblock {\em Waves in Random Media}, 2(3):209--224, 1992.

\bibitem{Linhai2011}
Huang Linhai and Changhui Rao.
\newblock {Wavefront sensorless adaptive optics: a general model-based
  approach.}
\newblock {\em Opt. Express}, 19(1):371--379, 2011.

\bibitem{Noll76}
Robert~J. Noll.
\newblock {Zernike polynomials and atmospheric turbulence}.
\newblock {\em J. Opt. Soc. Am.}, 66(3):207, 1976.

\bibitem{Roddier1999}
Francois Roddier, editor.
\newblock {\em {Adaptive optics in astronomy}}.
\newblock Cambridge University Press, 1999.

\bibitem{Schwertner2004}
M~Schwertner, M~Booth, and T~Wilson.
\newblock {Characterizing specimen induced aberrations for high NA adaptive
  optical microscopy.}
\newblock {\em Opt. Express}, 12(26):6540--6552, 2004.

\bibitem{Schwertner2004-charachterisation}
M.~Schwertner, M.~J. Booth, and T.~Wilson.
\newblock {Simulation of specimen-induced aberrations for objects with
  spherical and cylindrical symmetry}.
\newblock {\em J. Microsc.}, 215(3):271--280, 2004.

\bibitem{oleg_soloviev_2017_801515}
Oleg Soloviev.
\newblock {Optimal modes for wavefront sensorless adaptive optics. Turbulence-
  and oocyte-induced phase screens and Mathematica notebook.}
\newblock \url{https://doi.org/10.5281/zenodo.1196018}, 2018.

\bibitem{Thayil2011}
Anisha Thayil and Martin~J. Booth.
\newblock {Self calibration of sensorless adaptive optical microscopes}.
\newblock {\em Journal of the European Optical Society: Rapid Publications},
  6:11045, 2011.

\bibitem{Trevino2013}
Juan~P. Trevino, Jesus~E. G{\'o}mez-Correa, D.~Robert Iskander, and Sabino
  Ch{\'a}vez-Cerda.
\newblock Zernike vs. {B}essel circular functions in visual optics.
\newblock {\em Ophthalmic Physiol. Opt.}, 33(4):394--402, 2013.

\bibitem{Turaga2010}
Diwakar Turaga and Timothy~E Holy.
\newblock {Image-based calibration of a deformable mirror in wide-field
  microscopy}.
\newblock {\em Appl. Opt.}, 49(11):2030, 2010.

\bibitem{Vdovin2008}
Gleb Vdovin, Oleg Soloviev, Alexander Samokhin, and Mikhail Loktev.
\newblock {Correction of low order aberrations using continuous deformable
  mirrors}.
\newblock {\em Opt. Express}, 16(5):2859, 2008.

\bibitem{Wang2009}
Biru Wang and Martin~J. Booth.
\newblock Optimum deformable mirror modes for sensorless adaptive optics.
\newblock {\em Opt. Commun.}, 282(23):4467 -- 4474, 2009.

\bibitem{Yang2015}
Huizhen Yang, Oleg Soloviev, and Michel Verhaegen.
\newblock {Model-based wavefront sensorless adaptive optics system for large
  aberrations and extended objects}.
\newblock {\em Opt. Express}, 23(19):24587, 2015.

\end{thebibliography}

%
%
%
%
%

\end{document}